\documentclass[aps,prd,reprint,superscriptaddress,%
amsfonts,amsmath,amssymb,showkeys,floatfix, nofootinbib]{revtex4-1}
\usepackage{siunitx}
\usepackage{graphicx}
\usepackage{xcolor}
\usepackage{hyperref}
\hypersetup{
	pdftitle={Quantum Corrected Black Holes from String T-Duality},
	pdfauthor={Piero Nicolini, Euro Spallucci, Michael F. Wondrak},
	pdffitwindow=true,
	colorlinks=true,
	linkcolor={red!50!black},
	citecolor={blue!50!black},
	urlcolor={blue!80!black}
}

\newcommand{\inm}[1]{\ensuremath{\text{#1}}}
\newcommand{\diag}{\ensuremath{ \text{diag} }}
\newcommand{\lz}{\ensuremath{ {l_\inm{0}} }}
\newcommand{\Ldual}{\ensuremath{ {L} }}
\newcommand{\EpMod}{\ensuremath{ \mathfrak{T} }}
\newcommand{\pr}{\ensuremath{ {p_\inm{r}} }}
\newcommand{\pt}{\ensuremath{ {p_\inm{t}} }}
\newcommand{\lpl}{\ensuremath{ {l_\inm{P}} }}
\newcommand{\mpl}{\ensuremath{ {m_\inm{P}} }}
\newcommand{\GN}{\ensuremath{ {G_\inm{N}} }}
\newcommand{\lambdaCompRed}{{\lambdabar_\inm{C}}}

\newcommand{\iu}{\ensuremath{ {\text{i}} }}
\newcommand{\piu}{\text{\ensuremath{ \pi }}}
\newcommand{\dext}{\ensuremath{ \inm{d} }}
\newcommand{\der}{\ensuremath{ d }}
\newcommand{\Laplace}{\ensuremath{ \Delta }}
\newcommand{\cov}{\ensuremath{ \nabla }}
\DeclareMathOperator{\arsinh}{arsinh}

\begin{document}

\title{Quantum Corrected Black Holes from String T-Duality}

\author{Piero Nicolini}
\email{nicolini@fias.uni-frankfurt.de}

\affiliation{%
Frankfurt Institute for Advanced Studies (FIAS), 
Ruth-Moufang-Stra\ss{}e~1, 60438 Frankfurt am Main, Germany%
}

\affiliation{%
Institut f\"{u}r Theoretische Physik, Johann Wolfgang
Goethe-Universit\"{a}t Frankfurt am Main, 
Max-von-Laue-Stra\ss{}e~1, 60438 Frankfurt am Main, Germany%
}

\author{Euro Spallucci}
\email{spallucci@ts.infn.it}

\affiliation{%
Dipartimento di Fisica, Universit\`a degli Studi di Trieste, 
Strada Costiera 11, 34151 Trieste, Italy%
}

\affiliation{%
Istituto Nazionale di Fisica Nucleare (INFN), Sezione di Trieste, 
Strada Costiera 11, 34151 Trieste, Italy%
}

\author{Michael F.~Wondrak}
\email{wondrak@fias.uni-frankfurt.de}

\affiliation{%
Frankfurt Institute for Advanced Studies (FIAS), 
Ruth-Moufang-Stra\ss{}e~1, 60438 Frankfurt am Main, Germany%
}

\affiliation{%
Institut f\"{u}r Theoretische Physik, Johann Wolfgang
Goethe-Universit\"{a}t Frankfurt am Main, 
Max-von-Laue-Stra\ss{}e~1, 60438 Frankfurt am Main, Germany%
}

\date{August 21, 2019}

\begin{abstract}
In this letter we present some stringy corrections to black hole spacetimes emerging from string T-duality. 
As a first step, we derive the static Newtonian potential by exploiting the relation between the T-duality and the path integral duality. We show that the intrinsic non-perturbative nature of stringy corrections introduce an ultraviolet cutoff known as zero-point length in the path integral duality literature. As a result, the static potential is found to be regular. We use this result to derive a consistent black hole metric for the spherically symmetric, electrically neutral case. It turns out that the new spacetime is regular and is formally equivalent to the Bardeen metric, apart from a different ultraviolet regulator. On the thermodynamics side, the Hawking temperature admits a maximum before a cooling down phase towards a  thermodynamically stable end of the black hole evaporation process. The findings support the idea of universality of quantum black holes.
\end{abstract}


\maketitle

\section{Introduction.}
\label{sec:Introduction}

General relativity (GR) is in excellent agreement with astronomical observations and experimental findings and has passed several high-precision tests \cite{Will2014}. There is, however, a fundamental problem:  the classical description of the gravitational field breaks down at  short scales. Curvature singularities plague GR and attempts to amend it by the formulation of a quantum theory of gravity  have not been completely satisfactory so far.
In particular it is still missing a derivation of a singularity free black hole or a cosmological spacetime  from the first principles of  a theory of quantum gravity.

Given this background, there have been attempts to amend the bad short distance behavior of black hole solutions by assuming, as a guiding principle, the boundedness of spacetime curvature. Being the gravitational coupling $\GN = \lpl^2$ in natural units, the curvature cannot exceed $\lpl^{-2}$, where $\lpl$ is the Planck length. Probably the first example of regular black hole is the Bardeen spacetime \cite{Bardeen1968}. By assuming a non-linear electrodynamics Lagrangian, the ultraviolet cut off is obtained in terms of the magnetic charge of the black hole itself \cite{Beato2000}. This work has been followed by further singularity free models based on non-linear electrodynamics \cite{AyG98,AyG99}. 
Magnetic monopoles are, however, not yet observed in nature. Furthermore, the electric charge is quickly shed via Hawking emission and Schwinger pair production mechanism \cite{Gib75,ANS07,Nic18}.

To overcome such a limitation, neutral black hole models with a regular central core have been proposed in the literature, by a process of ``black hole engineering'', \textit{i.e.}~by assuming a limiting curvature without a derivation from first principles of a given theory \cite{AKM87,APS89,FMM89,FMM90,Dym92,Hay06} (for a review see \cite{Ansoldi2008} and the references therein). 

A major breakthrough has been achieved with a family of regular black holes that have been derived in a string-inspired way by averaging noncommutative coordinates on suitable coherent states \cite{NicoliniSS2006,NiS10,Nicolini2009}. Apart from the singularity freedom, such black holes offer an intriguing scenario for the end stage of the evaporation. Rather than a divergent Hawking emission, one finds a SCRAM phase, namely a cooling down towards a zero temperature extremal configuration, even in the absence of charge and angular momentum \cite{NiW11}.  
Interestingly, noncommutative effects have later been found to be equivalent to a non-local deformation of the Einstein-Hilbert action \cite{ModestoMN2011,Nicolini2012}. This fact has propelled the recent interest about black hole solutions in non-local gravity \cite{IMN13,NicoliniS2014,Fro15,FZd15,FrassinoKN2016}. Further string-based solutions can be found for instance in \cite{Maldacena1996,Mathur2005,CanoCOR2019}.

In this letter we aim to do a further step forward. We propose to exploit the ultraviolet finiteness of string theory to tame the singularity of classical black holes. To reach the goal, we consider, as a key ingredient, the concept of T-duality. The latter identifies string theories on higher-dimensional spacetimes with mutually inverse compactification radii~$R$. Quantum numbers are transformed correspondingly.
In the case of one compact dimension, the replacement rules are  
$R \rightarrow {R^\star}^2/R$ and $n \leftrightarrow w$,
where $R^\star = \sqrt{\alpha'}$ denotes the self-dual radius, $\alpha'$ is the Regge slope, $n$ stands for the Kaluza-Klein excitation and $w$ for the winding number of the closed string mode.

Interestingly, a similar duality is enjoyed also by fields when they probe the quantum fluctuations of the background spacetime \cite{Padmanabhan1997,Padmanabhan1998,SrinivasanSP1998}. 
Normally, when calculating a field propagator in the Schwinger representation, the larger is the weight given to a path, the shorter is the length of the path itself. Shorter paths are more strongly affected by spacetime fluctuations and are expected to be suppressed in the path integral.
The largest weight should be given to ``mid-sized paths'' of length comparable with a fundamental length scale $L\sim \lpl$.
This suggests a duality invariance of the path integral that is realized by assigning the weight $\exp\!\left(-m\left(s +\Ldual^2/s\right)\right)$, namely equal weight to paths of proper time $s$ and those of proper time $\Ldual^2/s$. 
The final quantum-corrected propagator is regular. It shows an ultraviolet cutoff, the so-called zero-point length $\lz$, which is assumed to equal $2\Ldual$ \cite{Padmanabhan1998}.

Excitingly, it is possible to embed the idea of the path integral duality into the framework of string T-duality. 
In the context of bosonic string theory, one can consider closed strings in a $(3+1)+1$-dimensional spacetime \cite{SmailagicSP2003,SpallucciF2005,FontaniniSP2006}.
In order to make contact with the Schwinger proper time formalism, one has to focus on the string center-of-mass and derive its propagator in an effective $3+1$-dimensional spacetime. The path-integral dual propagator arises from the T-dual propagator when expanded to first order.
This allows for a crystalline identification of parameters between the T-dual string theory and the path integral dual field theory, namely the zero-point length turns out to be equal to 
$\lz = 2\piu R^\star = 2\piu \sqrt{\alpha'}$. The ultraviolet finiteness of both the above string/field formulations provides a road map to derive regular black hole solutions.

We start by deriving the quantum-corrected static interaction potential according to the field theory enjoying the path integral duality.
We focus on the case of a scalar interaction as only the kinetic modification, but not the tensor structure, is relevant.
As an important background result, we recall that the momentum space propagator induced by the path integral duality is given, in the massless case, by 
\begin{equation}
G(k)
= -\frac{\lz}{\sqrt{k^2}}\, K_1 \!\left(\lz \sqrt{k^2}\right)
\end{equation}
where $\lz$ denotes the zero-point length of spacetime and $K_1\!\left(x\right)$ is a modified Bessel function of the second kind  \cite{Padmanabhan1997,SmailagicSP2003,SpallucciF2005,FontaniniSP2006}. At small momenta, \textit{i.e.}~for ${(\lz k)}^2 \to 0$, we obtain the conventional massless propagator $G(k) = -k^{-2}$.
At large momenta, the exponential suppression is responsible for curing UV divergences,
$G(k)
\sim -\lz^{1/2}\, {\left(k^2\right)}^{-3/4}\, e^{-\lz \sqrt{k^2}}
$.
%
The potential corresponding to the virtual-particle exchange can be obtained from the path integral of the force-mediating field. We consider a static external source $J$ which consists of two pointlike masses, $m$ and $M$, at relative distance $\vec{r}$. The generating functional of connected diagrams, $W[J]$, equals the integral of the interaction energy over the time span $T$. The potential follows to be
\begin{align}
V(r)
&= -\frac{1}{m}\, \frac{W[J]}{T}\\
&= -M\, \int\!\frac{\dext^3 k}{{\left(2\pi\right)}^3}\; {\left.G_\inm{F}(k)\right|}_{k^0=0}\; 
 \exp\!\left(\iu \vec{k} \cdot \vec{r}\right)\\
&= -\frac{M}{\sqrt{r^2 + \lz^2}},
\label{eq:statpot}
\end{align}
induced by the path integral duality or, equivalently, by T-duality. To our best knowledge, this is a new result which has not been derived in previous literature.
In contrast to the ordinary $1/r$-divergence, quantum spacetime fluctuations naturally implement a short-distance cutoff which acts non-perturbatively.

\section{T-duality black hole.}
\label{sec:BH}
The above result for the potential can be interpreted as a smearing effect of the matter distribution in contrast to conventional pointlike sources. The energy density of the modified matter follows from the Poisson equation,
\begin{equation}
\rho(r)
= \frac{1}{4\pi}\Laplace\, V(r)
= \frac{3 \lz^2 M}{4 \pi {\left(r^2 +\lz^2\right)}^{5/2}}.
\label{eq:rho}
\end{equation}
This fact opens the route to consistently study the quantum modifications of the spacetime geometry. Much in the same way as for the Poisson equation, one can obtain quantum corrections  of the Einstein tensor by coupling gravity to a quantum corrected energy momentum tensor $\EpMod_{\mu\nu}$ \cite{NicoliniSS2006,ModestoMN2011}.

\begin{figure}[t]
	\includegraphics[width=\linewidth]{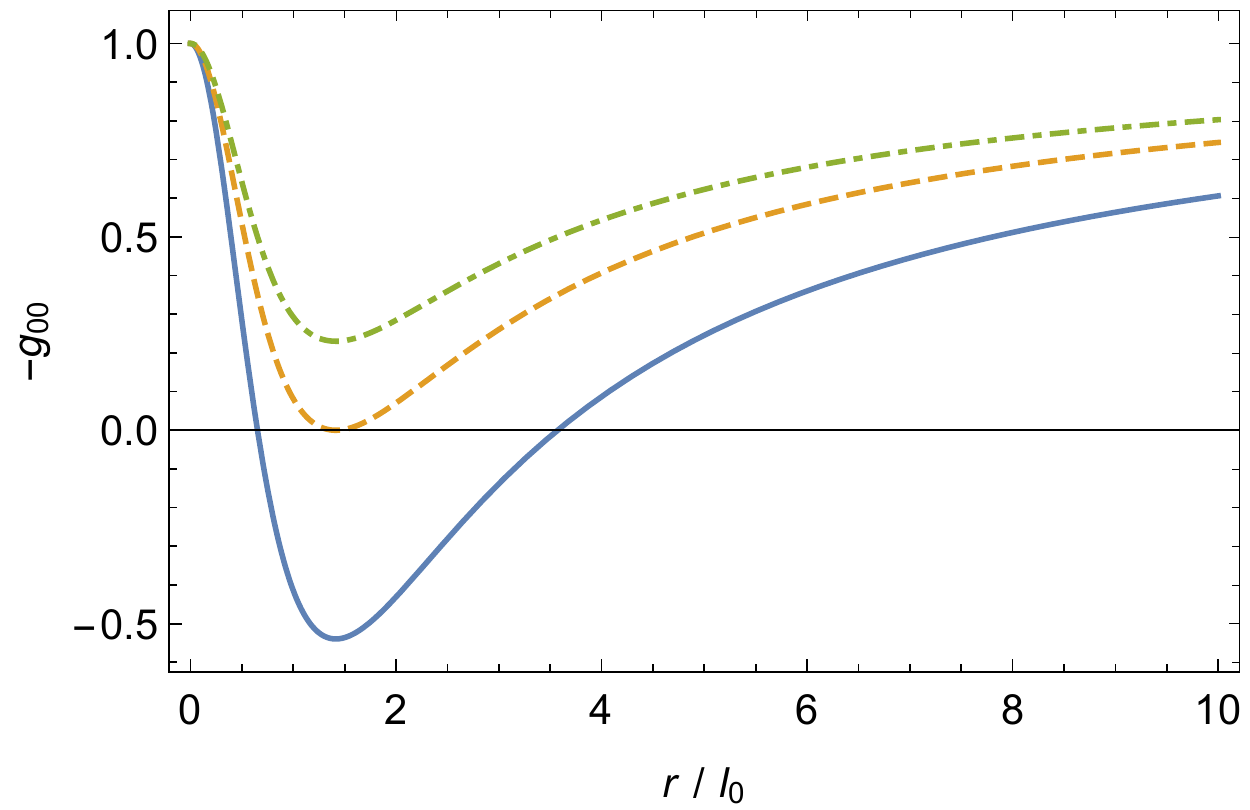}
	\caption{Metric coefficient $-g_{00}$ as a function of the radial distance $r$. The solid, dashed and dot-dashed lines respectively represent a black hole with mass $M = 2\lz > M^\inm{extr}$, an extremal black hole with mass $M = M^\inm{extr}$, and a regular horizonless geometry generated the the mass $M = \lz < M^\inm{extr}$.}
	\label{fig:2_Mg00-r}
\end{figure}

The general solution in case of a static, spherically symmetric source reads
\begin{equation}
\label{eq:lineElem}
\der s^2
= g_{00}\, \dext t^2 +g_{rr}\, \dext r^2 +r^2\, \dext \Omega^2
\end{equation}
with
\begin{equation}
\begin{split}
g_{00}(r)
&= -\left(1 - \frac{2 m(r)}{r}\right)
\end{split}
\end{equation}
and
$m(r)
= 4 \pi \int_0^r\! \dext \bar{r}\; \bar{r}^2\, \rho(\bar{r}).$
The Schwarzschild geometry corresponds to the case of a pointlike source, namely $\rho(\vec{x})=M\delta(\vec{x})$. Accordingly, the energy density, $\rho(r)$ in \eqref{eq:rho}, corresponds to the component $\EpMod_0{}^0$ of the quantum-modified energy-momentum tensor in static coordinates. Spherical symmetry leads to the following form in Schwarzschild-like coordinates: 
$(\EpMod_\mu{}^\nu) = \diag\left(-\rho,\pr,\pt,\pt\right)$. Here, $\pr$ and $\pt$ are the radial and transverse pressures, respectively. 
From energy-momentum conservation, $\cov_\mu \EpMod^{\mu\nu}=0$, and from the condition
$g_{rr} = -g_{00}^{-1}$ as in the Schwarzschild solution, the pressures are uniquely fixed in terms of the energy density. 
%
The only free metric degree of freedom is parametrized by the mass function $m(r)$ that after integration leads to the following metric coefficient
\begin{equation}
\label{eq:g00}
\begin{split}
-g_{00}(r)=
g_{rr}^{-1}(r)&
= 1 - \frac{2 M r^2}{{\left(r^2 +\lz^2\right)}^{3/2}}
\end{split}
\end{equation}
with $M$  the Komar mass (see Fig.~\ref{fig:2_Mg00-r}). 

The first surprising feature of the above spacetime geometry is that it is equivalent to the Bardeen black hole after changing the ultraviolet cutoff, namely by replacing the magnetic charge with  $\lz$. This means that the spacetime is regular! Stringy effects produce a de Sitter core at the origin, \textit{i.e.}~$g_{00} \simeq -1 +\Lambda_\inm{eff}\, r^2/3$ for $r \ll \lz$, with an effective cosmological constant $\Lambda_\inm{eff} = 6M/\lz^3$. This repulsive effect on small distances stabilizes the matter configuration against collapse. We will refer to  \eqref{eq:g00} as the ``T-duality black hole''.

The limit $\lz\to 0$ provides a cross check for consistency of our calculations. Throughout the letter, the ordinary GR results are recovered. Furthermore, the (infinite) series expansion in eq.~\eqref{eq:g00} demonstrates that $\lz$ appears in a non-perturbative way. It is only due to such a non-perturbative nature of the quantum corrections that the central singularity in the Schwarzschild case is tamed.

Another novelty of the solution is its mirror symmetry. Being the spacetime regular, it is geodesically complete. However in contrast to regular black holes obtained in the context of noncommutative geometry \cite{NicoliniSS2006}, the transformation $r\to -r$ does not generate negative mass solutions but a replica of the same universe \cite{MaN11}. Only the transformation $M\to -|M|$ generates another universe, \textit{i.e.}~a regular soliton of negative mass. We recall that regions of negative energy density might occur in the early universe for the quantum fluctuation at short scales \cite{MTY88,Man97}. As a result, both positive and negative mass solutions should be considered as  legitimate instantons and are expected to modify the standard  scenario  of de Sitter space decay \cite{MaRoss95,BoH96}.

The positive real roots of the metric component $g_{00}$ correspond to the horizon positions relative to the mass distribution. For $M> \frac{3 \sqrt{3}}{4}\, \lz$ there exist two roots, the inner and outer horizon,  $r_{-}$ and $r_{+}$, respectively. 
For large black hole masses $M$, the outer (event) horizon $r_{+}$ tends to the Schwarzschild value
\begin{equation}
r_{+}
= 2M -\frac{3 \lz^2}{4 M}  
+\mathcal{O}\!\left( \frac{\lz^3}{M^3} \right)
\end{equation}
up to a $1/M$ correction that resembles the $1/N$ quantum hair of the black hole quantum $N$ portrait \cite{DvG13,DvG13b,DvG14} as well as  what was recently found in the context of the generalized uncertainty principle \cite{CMN15}. On the other hand, in such a limit, the inner horizon approaches the origin 
\begin{equation}
r_{-}
= \frac{\lz}{\sqrt{2}}\, {\left(\frac{\lz}{M}\right)}^{1/2} 
+\mathcal{O}\!\left( \frac{\lz^{3/2}}{M^{3/2}} \right).
\end{equation}

For $M=M^\inm{extr} \equiv \frac{3 \sqrt{3}}{4}\, \lz \approx 1.30\, \lz$ there exists an extremal configuration with horizon radius $r_{-} =r_{+} =r^\inm{extr} =\sqrt{2}\, \lz \approx 1.41\, \lz$.
For smaller values of $M$, there is no root of $g_{00}$ and the geometry is a horizonless regular spacetime. 

The extremal configuration is of particular interest because it corresponds to a thermodynamically stable remnant configuration as will be shown below.
More importantly, the extremal configuration plays a fundamental role in the so called gravity self-completeness paradigm \cite{DvaliG2010,DvaliFG2011,SpA11,MuN12,Nic18}. Normally in particle collisions matter is compressed to length scales of the order of the Compton wavelength $\lambdaCompRed$. However, if a particle is accelerated to an energy larger than $M^\inm{extr}$, it is compressed below its corresponding horizon. This means that the particle undergoes a gravitational collapse into a black hole. In such a context the extremal configuration, being the smallest size black hole, is the natural candidate for the particle-black hole configuration. 
The latter separates the two possible phases of matter, the particle sector ($M < M^\inm{extr}$) and the black hole sector ($M > M^\inm{extr}$).
From the condition
$r^\inm{extr}
\stackrel{!}{=} \lambdaCompRed\!\left(M^\inm{extr}\right)$, we directly obtain:
\begin{align}
\label{eq:PhysVallz}
\lz 
&= \frac{2^{3/4}}{3^{3/4}}\, \lpl
\approx \num{0.738}\, \lpl\\
M^\inm{extr}
&= \frac{3^{3/4}}{2^{5/4}}\, \mpl
\approx \num{0.958}\, \mpl\\
r^\inm{extr}
&= \frac{2^{5/4}}{3^{3/4}}\, \lpl
\approx \num{1.04}\, \lpl
\end{align}
Eq.~\eqref{eq:PhysVallz} determines the value of the zero-point length. In the context of the path integral duality, we confirm $\lz \sim \lpl$ as originally assumed in \cite{Padmanabhan1997}. Moreover, we can also provide the Regge slope with a physical value,
\begin{equation}
\sqrt{\alpha'}
= \frac{\lz}{2\pi}
= \frac{1}{2^{1/4}\, 3^{3/4}\, \piu}\, \lpl
\approx \num{0.117}\, \lpl.
\end{equation}

\section{Black Hole Thermodynamics.}
\label{sec:Thermodynamics}
\begin{figure}[t]
	\includegraphics[width=\linewidth]{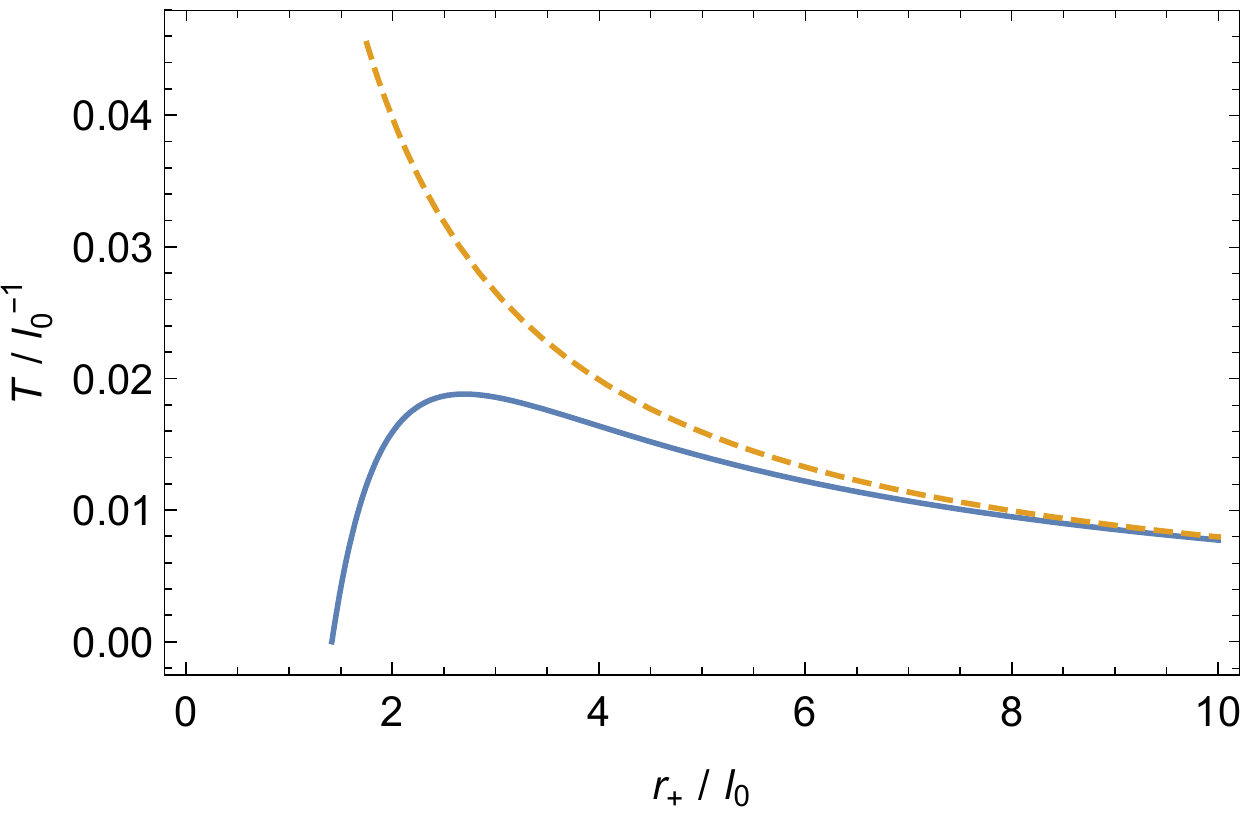}
	\caption{Hawking temperature $T$ 
	for the T-duality (solid) and Schwarzschild (dashed) black holes. 
	}
	\label{fig:4_T-r+ASS}
\end{figure}

\begin{figure}[t]
	\includegraphics[width=\linewidth]{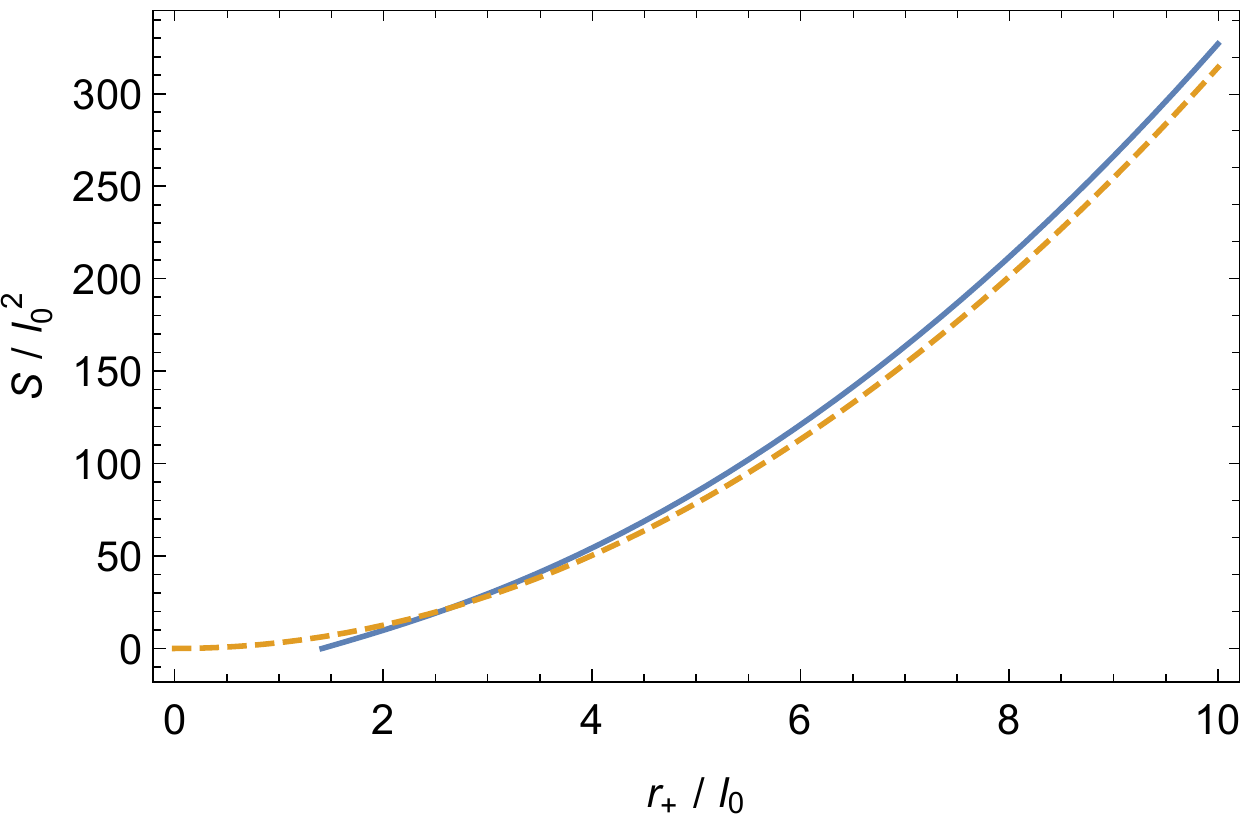}
	\caption{Black hole entropy $S$ as a function of the black hole radius $r_{+}$ for T-duality (solid) and Schwarzschild (dashed) black holes. By construction, the entropy of T-duality black holes becomes zero at $r_{+} = r^\inm{extr}$. At large distances the curves match up to logarithmic corrections one can find also in other quantum gravity proposals \cite{CR1996,Solo1997,Robb1998,Steve2000,Kaul2000}.}
	\label{fig:5_S-r+ASS}
\end{figure}

From \eqref{eq:g00} the Hawking temperature of the black hole reads
\begin{equation}
T
= {\left.\frac{\left|g_{00}'\right|}{4\pi}\right|}_{r=r_{+}}
= \frac{1}{4\pi\,r_{+}}\, \left(1 -\frac{3 \lz^2}{r_{+}^2 +\lz^2}\right)
\end{equation}
and is presented in Fig.~\ref{fig:4_T-r+ASS}. It has a maximum at
$r^\inm{crit} 
= \sqrt{\frac{7 +\sqrt{57}}{2}}\, \lz.
$
The black hole entropy, $S$, is obtained by integrating the first law of black hole thermodynamics in the form $\dext M = T\, \dext S$ with the boundary condition 
$S(r^\inm{extr}) = 0$:
\begin{align}
S(r_{+})
\label{eq:Entropy}
 &= \frac{A_{+}}{4} \left( 
  \left(1 -\frac{8\piu\lz^2}{A_{+}}\right) \sqrt{1 +\frac{4\piu\lz^2}{A_{+}}}\right.\\
 &\qquad \left.{}+\frac{12\piu\lz^2}{A_{+}} \left[\arsinh\sqrt{\frac{A_{+}}{4\piu\lz^2}} 
  -\arsinh\sqrt{2}\right]\vphantom{\sqrt{1 +\frac{\lz^2}{r_{+}^2}}}\right). \nonumber
\end{align}
Here, $A_{+} = 4\piu r_{+}^2$ denotes the horizon area. 
See Fig.~\ref{fig:5_S-r+ASS}.
%
%
Finally, we consider the black hole heat capacity that reads:
\begin{equation}
C 
= \frac{\der M}{\der T}
= -\frac{2\piu\, \left(r_{+}^2 -2\lz^2\right) {\left(r_{+}^2 +\lz^2\right)}^{5/2}}%
 {r_{+}^5 -7\lz^2\, r_{+}^3 -2 \lz^4\, r_{+}}.
\end{equation}
%

Equation~\eqref{eq:Entropy} 
explicitly demonstrate that the entropy is finite confirming the conjecture in ~\cite{Padmanabhan1998}. 
In general, T-duality black holes correctly reproduce the thermodynamic properties of the corresponding GR solution, \textit{i.e.}~Schwarzschild, for large distances, $r/\lz \gg 1$. For small distances, we find peculiar deviations. First of all, there are black holes for all values of $r_{+}$ in the case of Schwarzschild while only above the threshold $r^\inm{extr}$ for the T-dual metric. It is interesting to note that \eqref{eq:Entropy} has been obtain by an analytic integration of the first law of thermodynamics. Furthermore \eqref{eq:Entropy} is consistent with the area law 
\begin{equation}
dS=\frac{dM}{T}=\frac{dA_+}{4G(r_+)},
\end{equation}
provided one introduces an effective gravitational coupling
\begin{equation}
G(r)=G_\mathrm{N}\frac{r^3}{(r^2+l_0^2)^{3/2}},
\end{equation}
via the metric coefficient
\begin{equation}
-g_{00}=1-\frac{2G_\mathrm{N}M}{r}\frac{r^3}{(r^2+l_0^2)^{3/2}}\equiv 1-\frac{2G(r)M}{r}.
\end{equation}
The presence of an effective gravitational coupling is a feature present also in other models (see for instance \cite{Nic2010PRD}).

 For horizon radii in the range 
$r^\inm{extr} \leq r_{+} \leq r^\inm{crit}$, the temperature of the T-dual spacetime increases with the horizon size and hence with the mass (cf.~Fig.~\ref{fig:4_T-r+ASS}). This implies a positive heat capacity and thermodynamical stability. At $r^\inm{crit}$, the existence of a divergence of the heat capacity indicates a first-order phase transition between the phase of positive and that of negative heat capacity. 

In the context of Hawking evaporation, a black hole of size smaller than $r^\inm{crit}$ deviates from the Schwarzschild behavior. Instead of getting increasingly hotter with a final explosion, it cools down (SCRAM phase \cite{Nicolini2009}). The extremal configuration carries a significant role: It is of vanishing temperature and of vanishing entropy. Thus it can be regarded as a thermodynamically stable remnant of the black hole evaporation process. 

The properties discussed above are similar for other quantum-corrected black hole models, cf.~\textit{e.g.}~\cite{NicoliniSS2006,Nicolini2009,Nicolini2012,Mureika2019,NicoliniS2014,FrassinoKN2016,%
Mureika2008,GaeteHS2010,BonannoR2000,Modesto2006,ModestoP2009}. Therefore, the T-duality black hole promotes the notion of universal black hole characteristics in the quantum regime.

\section{Conclusions and Outlook.}
\label{sec:Conclusions&Outlook}
Despite string theory is a candidate for the quantum completion of gravity, there is yet no stringy description for black holes at full quantum level. In the presence of supersymmetry, D-branes show similar properties to classical black holes \cite{Polchinski1996,Maldacena1996}. In the context of the AdS/CFT correspondence, there are hints towards a microscopic explanation of the semiclassical Bekenstein entropy \cite{StromingerV1996}. The problem of the curvature singularity, however, is not solved. 

In contrast, the fuzzball concept is an exotic proposal which replaces the ordinary black hole by a condensed string state \cite{Mathur2005}. Avoiding a singularity at the center only works at the expense of sacrificing the very concept of an event horizon. Indeed, there is no classical limit in which one recovers the ordinary GR description.

The present letter contributes to fill this gap in the literature. We showed that it is possible to extract relevant string corrections to the classical description. 
%
%
%
%
We believe that our solution is opening a new route in quantum gravity, black holes and string theory that deserves further investigations in the near future.

\begin{acknowledgments}
MFW expresses his thanks to the Stiftung Polytechnische Gesellschaft Frankfurt am Main for their support. The work of PN has been supported by the grant NI 1282/3-1 of the project 
``Evaporation of microscopic black holes" of the German Research Foundation (DFG), 
by the Helmholtz International Center for FAIR within the framework of the LOEWE 
program (Landesoffensive zur Entwicklung Wissenschaftlich-\"{O}konomischer Exzellenz) launched by the State of Hesse and partially by GNFM, the Italian National Group for Mathematical Physics.
The authors are grateful to E. Ay\'on-Beato and T. Padmanabhan  for fruitful comments and references.
\end{acknowledgments}

\end{document}